\documentstyle[amssymb,aps,graphicx,epsf]{revtex}

\begin{document}
\title{A six dimensional analysis of Maxwell's Field Equations }
\author{Ana Laura Garc\'{i}a-Perciante$^{*\,\symbol{126}}$, Alfredo
Sandoval-Villalbazo $^{*}$, and L.S. Garc\'{i}a Col\'{i}n.$^{+}$}
\address{~}
\address{$^*$Departamento de Ciencias, Universidad\\
Iberoamericana, Col. Lomas de Santa Fe~01210, Mexico~D.F., Mexico\\
E-Mail: alfredo.sandoval@uia.mx}
\address{$^{\symbol{126}}$ Center for plasma theory and computation, University of
Wisconsin-Madison, 1500 Engineering Drive,  Madison, WI 53706, USA \\
E-Mail: algarper@cptc.wisc.edu\\
$^{+}$ Departamento de F\'{\i }sica,\\
Universidad Aut\'{o}noma Metropolitana-Iztapalapa, Av. Pur\'{\i }sima y\\
Michoac\'{a}n S/N 09340, Mexico~D.F., Mexico\\
E-Mail: lgcs@xanum.uam.mx}
\maketitle

\begin{abstract}
A framework based on an extension of Kaluza's original idea of using a five
dimensional space to unify gravity with electromagnetism is used to analyze
Maxwell\'{}s field equations. The extension consists in the use of a six
dimensional space in which all equations of electromagnetism may be obtained
using only Einstein's field equation. Two major advantages of this approach
to electromagnetism are discussed, a full symmetric derivation for the wave
equations for the potentials and a natural inclusion of magnetic monopoles
without using any argument based on singularities.
\end{abstract}

\section{Introduction}

In a recent paper \cite{GSG}, we have discussed the method proposed by
Kaluza in 1921 \cite{Kaluza} to derive Maxwell's equations for an
electromagnetic filed in vacuum by treating electric charges in space as
sources of curvature as mass does in Einstein's theory of general
relativity. In Kaluza's original paper, whereas the equations with sources
follow directly form the field equations, the homogeneous ones are derived
form a geometrical identity which happens to yield the correct results only
with a cartesian metric. For generalized coordinates, this last result
ceases to hold true. This was specifically illustrated using a Minkowsky
space with spherical symmetry. Moreover, we showed further how to overcome
the difficulty by introducing a 6 X 6 metric in which the sixth row and
columns contain additional potentials ($Z_{i\text{, }}i=1,2,3;\frac \eta c$%
). $g_{66}$ was taken equal to one and the elements $g_{56}$ and $g_{66}$
later proved to be related with two vectors $M^\mu $ and $Q^\mu ,$ both
irrotational, and the full symmetric expressions for Maxwell's equations in
terms of magnetic point charges were recovered. The existence of these
latter ones, although still controversial, arise in a complete natural way
in the formalism.

It was also mentioned in paper \cite{GSG} that with the results obtained
therein, one can also arrive, correspondingly, to a set of completely
symmetric wave equations for the field potentials, thus removing the old
difficulty of being able to express the magnetic field as the curl of a
vector potential.

Indeed, if magnetic monopoles are accounted for, the divergence of the
magnetic field cannot vanish. This equation has led to different approaches
starting from the pioneering work of Dirac \cite{Dirac} who proposed the
idea of considering a singularity along an axis with the monopole at the
origin, so the magnetic field can still be taken as a curl in the vector
potential in every region of space not containing such line. This idea, as
well as other approaches have been discussed by many other authors \cite
{Zeleny} , all of them foreign to Kaluza's ideas.

What we explicitly want to show in this paper is that in the 6D formalism
presented in Ref. \cite{GSG}, the two additional potentials lead to two
equivalent formulations for the electromagentic field, and therefore, such
potentials can be treated in a completely symmetrical way. It is worth
noticing that a similar four potential has been considered before \cite
{Singleton}, but the expressions for the electromagentic fields differ from
those obtained within Kaluza's framework.

\smallskip We also want to stress that the inclusion of magnetic monopoles
within Kaluza's scheme is not new. It has been done by several authors in
different contexts using five dimensions and associating singularities with
monopoles \cite{Kaluza2}. What is entirely new in this treatment is the use
of a 6D metric which allows the introduction of monopoles keeping all
symmetries between the fields and the potentials.

In section II we briefly recapitulate the results of Ref. \cite{GSG}, in
section III we discuss the wave equations for the additional potentials and
derive the continuity equations for electric and magnetic charges. We
finally leave section IV for some pertinent concluding remarks.

\section{General Background}

As it has been discussed at length in the literature, physicists have long
searched for magnetic monopoles. The main reason is very simple: {\em %
symmetry}. Electromagnetic theory would become much more symmetric if terms
proportional to magnetic charges and current densities are added to the
Maxwell equations. This has been very clearly pointed out in the literature,
such equations adopt a very elegant and symmetric structure. But some
inconsistences appear if one turns to work with the electromagnetic
potentials. The magnetic field cannot be represented as a curl anymore,
since its divergence is now supposed to be proportional to a magnetic charge
density. In Jackson\'{}s textbook \cite{Jackson}, a solution {\em for a
particular problem} is mathematically stated without major obstacles. Some
other authors, in the spirit of Jackson, have proposed to represent a
magnetic monopole as a singularity in a four dimensional space-time in which
the vector potential is discontinuous. Since in this case the potential is
not differentiable, the definition of the field tensor, in terms of
potentials, ceases to hold. Then, the magnetic field would have a vanishing
divergence in all the four dimensional space, except at the monopole\'{}s
position. This argument solves the problem but introduces an undesirable
asymmetry in the definitions of magnetic and electric potentials. An
alternative solution to this problem can be found following the line of
Kaluza\'{}s theory in a 6D space time. In order to include a magnetic charge
and preserve full symmetry in electromagnetic theory, five dimensions are
not enough. The fifth dimension introduced by Kaluza was characterized by
the condition $\frac{dx^5}{dt}=\frac qm$, $q$ being the charge, $\;m$ its
mass and the cylindrical condition $\frac \partial {\partial x^5}=0$.
Analagously, a sixth dimension is now introduced according to the conditions 
\begin{equation}
\frac{dx^6}{dt}=\frac gm  \label{mm1}
\end{equation}
\begin{equation}
\frac \partial {\partial x^6}=0  \label{mm2}
\end{equation}
where in Eq. (\ref{mm1}), $g$ is the magnetic charge and $m$ is the rest
mass of the particle.

For the sake of simplicity, the metric tensor to work with will consist in a
6D generalization of Minkowky\'{}s space time with the new entries: 
\begin{equation}
g_{n6}=g_{6n}=Z_n\;\;\;n=1,2,3  \label{mm3}
\end{equation}
and 
\begin{equation}
g_{46}=g_{64}=\frac \eta c  \label{mm4}
\end{equation}
Here, the quantities $Z_n$ and $\eta $ will be interpreted later. At present
we shall regard them as two potentials, one of scalar type $\eta $ and one
of a vector type $Z_n$. This is done in order to preserve symmetry with the
conventional theory. For the time being, the coefficients $g_{56}=g_{65}$
are left undetermined, but will be accounted in a rigorous way later. With
the proposed metric, the Chistoffel symbols become, 
\begin{equation}
\Gamma _{\beta 5}^\alpha =A_{,\beta }^\alpha -A_{,\alpha }^\beta +Z^\alpha
g_{56,\beta },  \label{mm5}
\end{equation}

\smallskip and 
\begin{equation}
\Gamma _{\beta 6}^\alpha =Z_{,\beta }^\alpha -Z_{,\alpha }^\beta +A^\alpha
g_{56,\beta }.  \label{mm6}
\end{equation}
where $A^n$ is the ordinary vector potential. The Christoffel symbols may
also be obtained by requiring that a magnetic field and an electrically
charged particle, as well, move along geodesics in this space-time, and
comparing its equation of motion with the Lorentz's force, which now
includes a symmetric term depending on $g$. An example of this procedure was
worked out by Stephani \cite{Stephani} in a pure gravitational context.
Thus, the proposed Lorentz force reads:

\begin{equation}
\frac{d^2x^\alpha }{dt^2}=\frac qm\left[ \varepsilon _{\beta \gamma }^\alpha 
\frac{\partial x^\beta }{\partial t}B^\gamma +E^\alpha \right] +\frac gm%
\left[ \varepsilon _{\beta \gamma }^\alpha \frac{\partial x^\beta }{\partial
t}E^\gamma -B^\alpha \right] ,  \label{Lorentz}
\end{equation}
where $\varepsilon _{\beta \gamma }^\alpha $ is Levi Civita's tensor. The
Christoffel symbols become those of the five dimensional theory, plus the
new ones corresponding to the sixth dimension namely,

\[
\Gamma _{6n}^{m}=\varepsilon _{sn}^{m}E^{s}, 
\]

\[
\Gamma _{64}^{n}=-B^{n}. 
\]

Comparing both sets of values for the Christoffel symbols (see Appendix A),
the definitions of the electric and magnetic fields are obtained. Because
there are two sets of Christoffel symbols, we obtain two equivalent forms of
expressing the electromagnetic field namely,

\begin{equation}
E^k=-\varphi ^{,k}-A_{,4}^k,  \label{Ecompd}
\end{equation}
\begin{equation}
B^k=\eta ^{,k}-Z_{,4}^k,  \label{Bcompd}
\end{equation}
\begin{equation}
E^k=\varepsilon _{lm}^kZ^{l,m}+M^k,  \label{Ecompr}
\end{equation}
\begin{equation}
B^k=\varepsilon _{lm}^kA^k+Q^k.  \label{Bcompr}
\end{equation}
If the vectors $M^k$, $Q^k$ are defined as: 
\begin{equation}
M^k=\left[ 
\begin{array}{l}
Z^2g_{65,3} \\ 
Z^3g_{65,1} \\ 
Z^1g_{65,2}
\end{array}
\right] \qquad Q^k=\left[ 
\begin{array}{l}
A^2g_{56,3} \\ 
A^3g_{56,1} \\ 
A^1g_{56,2}
\end{array}
\right] ~~~,  \label{MyQ}
\end{equation}
both fields can be represented in two different ways, and become completely
symmetric. Here $Z^n$ and $A^n$ with n=1,2,3 are the electric and magnetic
vector potentials respectively. The new set of definitions account for a
duality in the components of the electromagnetic field. With this dual
conception of the field tensor, all the theory can be reformulated including
magnetic monopoles and wave equations for the potentials can be obtained.
This was shown in Ref. \cite{GSG}.

Eqs. (\ref{Ecompr}) and (\ref{Bcompr}) show the importance of the quantities 
$g_{56}$ and $g_{65}$ which are postulated to be independent of time and
equal to each other (the metric tensor is symmetric). Then, introducing
expressions (\ref{Ecompr}) and (\ref{Bcompr}) in Maxwell's equations for the
divergence of the fields (Eqs. (\ref{MdE}) and (\ref{MdB}) ), the relation
between the new vectors $M^k$ and $Q^k$ is given by:

\begin{equation}
\nabla \cdot {\bf Q}=\frac{\rho }{\varepsilon _{0}}\frac{q}{m}~,
\label{divq}
\end{equation}

\begin{equation}
\nabla \cdot {\bf M}=\frac \rho {\varepsilon _0}\frac gm~,  \label{divm}
\end{equation}

The vectors ${\bf M}$ and ${\bf Q}$ in this equations contain space
derivatives of $g_{56}=g_{65}.$ If this quantity is proposed as a constant
value, then {\em both} electric and magnetic fields would have zero
divergence. In such a case, neither magnetic {\em nor electric} charges
would fit in the theory. Note that these vectors are irrotational, a fact
that is sustained by a generalized Helmholtz theorem \cite{Kobe}.

The role of these vectors becomes transparent when the new definitions for
the fields are replaced in the electromagnetic vector equations. The full
implications of this step will be discussed in the following section.

As it was mentioned in the introduction, there is an inherent asymmetry in
the way used by Kaluza to derive Maxwell\'{}s equations. The inhomogeneous
equations are derived from the field equation . The other two equations are
a consequence of making the cylindrical condition act over an expression\
that is identically zero by itself {\em so no additional information can be
extracted from it.} Instead, when the theory is formulated in 6 dimensions, 
{\em the field equation can be used twice, }once for each extra dimension
and therefore both pairs of equations can be obtained. In order to analyze
this last statement, we start from a stress tensor, which reads:

\begin{equation}
T^{5\nu }=\rho v^5v^\nu ,  \label{esfuerzos}
\end{equation}

\begin{equation}
T^{6\nu }=\rho v^6v^\nu .  \label{esf6}
\end{equation}
Here $\rho $ is the rest mass density and $v^\nu =\frac{dx^\nu }{ds}$ with $%
\mu ,$ $\nu $ running from one to six . Ricci's tensor is calculated by
contracting the Riemann-Christoffel curvature tensor and, considering the
curvature scalar equal to zero, the field equation becomes

\begin{equation}
G_{\mu \nu }=R_{\mu \nu }=\kappa \,T_{\mu \nu }  \label{fieldeq}
\end{equation}

The stress tensor includes now several components porportional to the
electric and magnetic current densities. Using $v^5=\frac qm$ and $v^6=\frac %
gm$ we have that,

\begin{equation}
R_{\mu 5}=\kappa \,T_{\mu }=\alpha J_{\mu }  \label{field5}
\end{equation}

\begin{equation}
R_{\mu 6}=\kappa \,T_{\mu 6}=\alpha K_{\mu }  \label{field6}
\end{equation}
where $J^{\mu }=\rho \frac{q}{m}v^{\mu }$ and $K^{\mu }=\rho \frac{g}{m}%
v^{\mu }$ with $\mu =1,2,3,4$.

Making $\mu =1,2,3,\,$expressions (\ref{field5}) and (\ref{field6}) become,
in vector form, 
\begin{equation}
\nabla \times {\bf E}=-\frac{1}{c}\frac{\partial {\bf B}}{\partial t}-{\bf K}
\label{MrE}
\end{equation}

\begin{equation}
\nabla \times {\bf B}=\frac{\partial {\bf E}}{\partial t}+\mu _{0}{\bf J}
\label{MrB}
\end{equation}
and if we take $\mu =4,$in those same equations, the following well-known
expressions are obtained:

\begin{equation}
\triangledown \cdot {\bf E}=\frac \rho {\varepsilon _0}\frac qm  \label{MdE}
\end{equation}

\begin{equation}
\triangledown \cdot {\bf B}=\mu _{0}\rho \frac{g}{m}  \label{MdB}
\end{equation}

In the absence of magnetic charge ($g=0$), the usual electromagnetic
equations are recovered. In the presence of magnetic charges, the stress
tensor has new components which, by means of the field equation, lead to a
full symmetric set of electromagnetic equations.

\section{Wave Equations for Electromagnetic Potentials}

After the lengthy but necessary account of the main results required to
tackle the crucial question in this paper, we will now assess the usefulness
of the 6D space in deriving the wave equation for the new potentials $\eta $
and $Z^n$ introduced in Eqs. (\ref{mm3}) and (\ref{mm4}). without resorting
to arguments concerning singularities. In fact, if we now substitute the
definitions of the vectors $E^k$, $B^k$ given in Eqs. (\ref{Ecompd}-\ref
{Bcompd}) into Eqs. (\ref{MdE}-\ref{MdB}), one finds that

\begin{equation}
\nabla ^2\eta +\frac \partial {\partial t}\nabla \cdot {\bf Z}=\mu _0\rho 
\frac gm.  \label{On2}
\end{equation}

It is important to notice that in order to obtain the wave equations, it
becomes necessary to introduce one additional fact, namely, the condition
that the Lorentz's gauge $A_{,\mu }^\mu =0$ has to be also applied to the
new vector potential $Z^n$, so that $Z_{,\mu }^\mu =0$ .

This condition implies

\begin{equation}
\nabla \cdot {\bf Z}+\frac 1{c^2}\frac{\partial \eta }{\partial t}=0,
\label{normZ}
\end{equation}
which, when replaced in eq. (\ref{On2}), gives

\begin{equation}
\nabla ^2\eta -\frac 1{c^2}\frac{\partial ^2\eta }{\partial t^2}=\mu _0\rho 
\frac gm.  \label{On}
\end{equation}
On the other hand, the alternative definition for ${\bf B}$ given in eq. (%
\ref{Bcompr}) and the usual one for ${\bf E}$ (\ref{Ecompd}) can be replaced
in equation (\ref{MrB}), so that

\begin{equation}
\nabla \times (\nabla \times {\bf A}+{\bf Q})=\mu _{0}{\bf J}+\mu
_{0}\varepsilon _{0}\frac{\partial }{\partial t}(-\triangledown \varphi -%
\frac{\partial {\bf A}}{\partial t}),  \label{OA1}
\end{equation}
which is easily reduced to the result that

\begin{equation}
\nabla (\nabla \cdot {\bf A})-\nabla ^{2}{\bf A}+\nabla \times {\bf Q}=\mu
_{0}{\bf J}-\frac{1}{c^{2}}\nabla \frac{\partial \varphi }{\partial t}-\frac{%
1}{c^{2}}\frac{\partial ^{2}{\bf A}}{\partial t^{2}}.  \label{OA2}
\end{equation}
Recalling that ${\bf Q}$ is irrotational (see section 4) and using the gauge
condition $A_{,\mu }^{\mu }=0$ we can rewrite Eq. (\ref{OA2}) as:

\begin{equation}
\nabla ^2{\bf A}-\frac 1{c^2}\frac{\partial ^2{\bf A}}{\partial t^2}=\mu _0%
{\bf J.}  \label{OA}
\end{equation}
Following an analogous procedure, the equation for ${\bf Z}${\bf \ }can also
be obtained. The result reads:

\begin{equation}
\nabla ^2{\bf Z}-\frac 1{c^2}\frac{\partial ^2{\bf Z}}{\partial t^2}=\frac 1{%
\varepsilon _0}{\bf K.}  \label{OZ}
\end{equation}
It is now important to remark that the above procedure could be questioned
by inquiring what would happen if the complementary definitions for ${\bf E}$
and ${\bf B}$ are introduced in the same field equation (\ref{MrE}) or (\ref
{MrB}). To examine this puzzle let us now take ${\bf E}$ defined by Eq. (\ref
{Ecompd}) and ${\bf B}$ by Eq. (\ref{Bcompr}) and substitute them into Eq. (%
\ref{MrE}). This will give rise to the following result,

\begin{equation}
\nabla \times (-\triangledown \varphi -\frac{\partial {\bf A}}{\partial t})=-%
\frac \partial {\partial t}(\triangledown \times {\bf A}+{\bf Q})-\frac 1{%
\varepsilon _0}{\bf K,}  \label{C1}
\end{equation}
and therefore

\begin{equation}
\frac{\partial {\bf Q}}{\partial t}+\frac{1}{\varepsilon _{0}}{\bf K}=0.
\label{C2}
\end{equation}
Moreover, introducing the definition given by eq. (\ref{Bcompr}) for ${\bf B}
$ in equation (\ref{MdB}) and taking divergence on both sides of it, the
following expressions are obtained:

\begin{equation}
\nabla \cdot (\nabla \times {\bf A}+{\bf Q})=\mu _{0}\rho \frac{g}{m},
\label{C3}
\end{equation}
or,

\begin{equation}
\nabla \cdot {\bf Q}=\mu _{0}\rho \frac{g}{m}.  \label{C4}
\end{equation}

Taking the time derivative of (\ref{C4}), the divergence in (\ref{C2}) and
combining the resulting equations, a continuity equation for the magnetic
charge is obtained, namely: 
\begin{equation}
\nabla \cdot \frac{\partial {\bf Q}}{\partial t}=\mu _0\frac{\partial (\rho 
\frac gm)}{\partial t},  \label{C5}
\end{equation}
or, in a more familiar notation,

\begin{equation}
\frac 1{c^2}\frac{\partial (\rho \frac gm)}{\partial t}+\nabla \cdot {\bf K}%
=0.  \label{C6}
\end{equation}

Therefore, such a procedure generates the continuity equations, both for
magnetic as well as electric charges in terms of their respective fluxes, $%
{\bf K}$ and ${\bf J}$. Thus the symmetry sought for the electromagnetic
field equations is complete.

Summarizing, in this section, wave equations for potentials have been
obtained. These equations are completely symmetric. When the new definitions
are introduced into Maxwell's equations, wave equations for potentials or
continuity equations for the charges are obtained.

\section{Concluding Remarks}

Under the assumption that assures the existence of magnetic monopoles,
working in a six-dimensional space-time gives useful information about the
nature of electromagnetic fields and potentials. In the traditional theory,
the expressions for the electric and magnetic field in terms of potentials
are different. This difference is caused by the absence of a magnetic charge
in the formalism and leads to an asymmetric set of Maxwell's equations. By
means of an extension of Minkowski's metric to six dimensions, symmetric
Maxwell's equations result from Einstein's field equation. As part of the
geometrical description inherent to this six dimensional space-time, an
electric vector potential and a magnetic scalar potential are required. With
this set of functions, {\em dual definitions for the fields can be formulated%
}, and, with them, four wave equations, one for each potential, $\eta $, $%
\varphi $, ${\bf A}$ and ${\bf Z}$ are obtained.

Magnetic monopoles are theoretically possible but none has been detected so
far. Living in a big universe induces people to think that the possibility
of its existence is not remote. Research is being done in this line.
Therefore, since no magnetic charges are found yet, it is not possible to
estimate the order of the charge-mass ratio of a magnetic monopole, without
using quantum theory. This ratio shall characterize the sixth dimension by
being equal to it's time variation, as is expressed by Eq.(\ref{mm1}).
Meanwhile, the assumption associating a small value for $g/m$ seems to be
correct, but this hypothesis certainly needs a stronger support.

This paper establishes something more than a different formulation of
results already obtained in magnetic monopole physics. The theory of a six
dimensional space-time gives additional information on the subject. The
duality of the field tensor has been mentioned before \cite{Kobe} but in
this work a natural origin of this property is exhibited. Six dimensions
implies two four-potentials, and so, two different symmetric definitions for
each field. Therefore, an electromagnetic theory including magnetic
monopoles, formulated in a six-dimensional space-time, gets closer to the
challenge of a universal description of all phenomena in nature based in the
geometry of space-time.

\bigskip \appendix

\section{Christoffel symbols and its relation with the electromagnetic field}

There are two different mechanisms by which Christoffel symbols can be
obtained. The first one consists in comparing the equation for a geodesic in
a six dimensional space with a corresponding equation of motion for a
charged particle under the influence of a generalized Lorentz's force. This
equation of motion reads

\begin{equation}
\frac{d^2x^\alpha }{dt^2}=\frac qm\left[ \varepsilon _{\beta \gamma }^\alpha 
\frac{\partial x^\beta }{\partial t}B^\gamma +E^\alpha \right] +\frac gm%
\left[ \varepsilon _{\beta \gamma }^\alpha \frac{\partial x^\beta }{\partial
t}E^\gamma -B^\alpha \right] .\medskip  \label{movgen}
\end{equation}
\medskip For example, let's take the first component of equation (\ref
{movgen}):

\begin{equation}
\frac{d^{2}x^{1}}{dt^{2}}=\frac{q}{m}\frac{\partial x^{2}}{\partial t}B_{3}-%
\frac{q}{m}\frac{\partial x^{3}}{\partial t}B_{2}+\frac{q}{m}E_{1}+\frac{g}{m%
}\frac{\partial x^{2}}{\partial t}E_{3}-\frac{g}{m}\frac{\partial x^{3}}{%
\partial t}E_{2}-\frac{g}{m}B_{1},  \label{movcomp1}
\end{equation}
which shall be compared, as mentioned before, with the first component of a
geodesic, namely,

\begin{equation}
\frac{d^{2}x^{1}}{dt^{2}}+\Gamma _{5\nu }^{1}\frac{q}{m}\frac{dx^{\nu }}{ds}%
+\Gamma _{6\nu }^{1}\frac{g}{m}\frac{dx^{\nu }}{ds}=0.\medskip
\label{geod61}
\end{equation}

Equation (\ref{geod61}), after running the indices from 1 to 6, turns out as
follows

\[
\frac{d^{2}x^{1}}{dt^{2}}+\Gamma _{51}^{1}\frac{q}{m}\frac{dx^{1}}{ds}%
+\Gamma _{52}^{1}\frac{q}{m}\frac{dx^{2}}{ds}+\Gamma _{53}^{1}\frac{q}{m}%
\frac{dx^{3}}{ds}+\Gamma _{54}^{1}\frac{q}{m}\frac{dx^{4}}{ds}+\medskip 
\]
\begin{equation}
\;\;\;\Gamma _{61}^{1}\frac{g}{m}\frac{dx^{1}}{ds}+\Gamma _{62}^{1}\frac{g}{m%
}\frac{dx^{2}}{ds}+\Gamma _{63}^{1}\frac{g}{m}\frac{dx^{3}}{ds}+\Gamma
_{64}^{1}\frac{g}{m}\frac{dx^{4}}{ds}=0\,.\medskip  \label{geod62}
\end{equation}
Now, comparing equations (\ref{movcomp1}) and (\ref{geod62}) one obtains

\begin{equation}
\Gamma _{51}^1=0,\smallskip   \label{c151}
\end{equation}
\begin{equation}
\Gamma _{52}^1=B_3,\smallskip   \label{c152}
\end{equation}
\begin{equation}
\Gamma _{53}^1=-B_2,\smallskip   \label{c153}
\end{equation}
\begin{equation}
\Gamma _{54}^1=\frac 1cE_1;\smallskip   \label{c154}
\end{equation}
and\smallskip 
\begin{equation}
\Gamma _{61}^1=0,\smallskip   \label{c161}
\end{equation}
\begin{equation}
\Gamma _{62}^1=E_3,\smallskip   \label{c162}
\end{equation}
\begin{equation}
\Gamma _{63}^1=-E_2,\smallskip   \label{c163}
\end{equation}
\begin{equation}
\Gamma _{64}^1=-\frac 1cB_1.\smallskip   \label{c164}
\end{equation}
The second mechanism to obtain Christoffel symbols consists in introducing
the metric elements in their definition, see ref. \cite{GSG}. The
expressions for symbols in equations (\ref{c151}) to (\ref{c164}) are given
below 
\begin{equation}
\Gamma _{51}^1=\frac 12g^{11}\left( \frac{\partial g_{51}}{\partial x^1}-%
\frac{\partial g_{51}}{\partial x^1}\right) +\frac 12g^{15}\frac{\partial
g_{55}}{\partial x^1}=0\bigskip ,  \label{h151}
\end{equation}
\[
\Gamma _{52}^1=\frac 12g^{11}\left( \frac{\partial g_{51}}{\partial x^2}-%
\frac{\partial g_{52}}{\partial x^1}\right) +\frac 12g^{16}\frac{\partial
g_{56}}{\partial x^2}\smallskip 
\]
\begin{equation}
=\frac 12\left( \frac{\partial A_1}{\partial x^2}-\frac{\partial A_2}{%
\partial x^1}\right) +\frac 12Z_1\frac{\partial g_{56}}{\partial x^2}%
,\bigskip   \label{h152}
\end{equation}
\[
\Gamma _{53}^1=\frac 12g^{11}\left( \frac{\partial g_{51}}{\partial x^3}-%
\frac{\partial g_{53}}{\partial x^1}\right) +\frac 12g^{16}\frac{\partial
g_{56}}{\partial x^3}\smallskip 
\]
\begin{equation}
=\frac 12\left( \frac{\partial A_1}{\partial x^2}-\frac{\partial A_3}{%
\partial x^1}\right) +\frac 12Z_1\frac{\partial g_{56}}{\partial x^3}%
,\bigskip   \label{h153}
\end{equation}
\begin{equation}
\Gamma _{54}^1=\frac 12g^{11}\left( \frac{\partial g_{51}}{\partial x^4}-%
\frac{\partial g_{54}}{\partial x^1}\right) +\frac 12g^{16}\frac{\partial
g_{56}}{\partial x^4}=\frac 12\left( \frac{\partial A_1}{\partial x^4}+\frac %
1c\frac{\partial \phi }{\partial x^1}\right) ;\bigskip   \label{h154}
\end{equation}
and 
\begin{equation}
\Gamma _{61}^1=\frac 12g^{11}\left( \frac{\partial g_{61}}{\partial x^1}-%
\frac{\partial g_{61}}{\partial x^1}\right) +\frac 12g^{16}\frac{\partial
g_{66}}{\partial x^1}=0\bigskip ,  \label{h161}
\end{equation}
\[
\Gamma _{62}^1=\frac 12g^{11}\left( \frac{\partial g_{61}}{\partial x^2}-%
\frac{\partial g_{62}}{\partial x^1}\right) +\frac 12g^{15}\frac{\partial
g_{65}}{\partial x^2}\smallskip 
\]
\begin{equation}
=\frac 12\left( \frac{\partial Z_1}{\partial x^2}-\frac{\partial Z_2}{%
\partial x^1}\right) +\frac 12A_1\frac{\partial g_{65}}{\partial x^2}%
,\bigskip   \label{h162}
\end{equation}
\[
\Gamma _{63}^1=\frac 12g^{11}\left( \frac{\partial g_{61}}{\partial x^3}-%
\frac{\partial g_{63}}{\partial x^1}\right) +\frac 12g^{15}\frac{\partial
g_{65}}{\partial x^3}\smallskip 
\]
\begin{equation}
=\frac 12\left( \frac{\partial Z_1}{\partial x^2}-\frac{\partial Z_3}{%
\partial x^1}\right) +\frac 12A_1\frac{\partial g_{65}}{\partial x^3}%
,\bigskip   \label{h163}
\end{equation}
\begin{equation}
\Gamma _{64}^1=\frac 12g^{11}\left( \frac{\partial g_{61}}{\partial x^4}-%
\frac{\partial g_{64}}{\partial x^1}\right) +\frac 12g^{15}\frac{\partial
g_{56}}{\partial x^4}=\frac 12\left( \frac{\partial Z_1}{\partial x^4}+\frac %
1c\frac{\partial \eta }{\partial x^1}\right) .\bigskip   \label{h164}
\end{equation}
The time independence of $g_{56}$ is justified by equations (\ref{h154}) and
(\ref{h164}). For example, the symbols calculated in Eq. (\ref{h154}) should
be $-E_1/c$, which, considering the traditional definition for the electric
field in terms of potentials should read $\left( \partial A_1/\partial
x^4+\partial (\phi /c)/\partial x^1\right) $.

Comparing equations (\ref{c151}) to (\ref{c164}) and (\ref{h151}) to (\ref
{h164}) the relations between the fields and the potentials can be expressed
as follows 
\begin{equation}
E^\nu =-\frac{\partial \phi }{\partial x^\nu }-\frac{\partial A_\nu }{%
\partial t},\smallskip   \label{Econv}
\end{equation}
\begin{equation}
E^\nu =\varepsilon _{\beta \gamma }^\nu \left( \frac{\partial Z_\beta }{%
\partial x^\gamma }-\frac{\partial Z_\gamma }{\partial x^\beta }\right)
+M^\nu ,\smallskip   \label{Enuev}
\end{equation}
\begin{equation}
B^\nu =-\frac{\partial \eta }{\partial x^\nu }-\frac{\partial Z_\nu }{%
\partial t},\smallskip   \label{Bnuev}
\end{equation}
\begin{equation}
B^\nu =\varepsilon _{\beta \gamma }^\nu \left( \frac{\partial A_\beta }{%
\partial x^\gamma }-\frac{\partial A_\gamma }{\partial x^\beta }\right)
+Q^\nu .\medskip   \label{Bconv}
\end{equation}
The reason for having two equivalent expressions for each field in terms of
potentials comes from a duality in Faraday's tensor which can be expressed
in terms of the equivalence in Christoffel symbols with indices 5 and 6. For
example, 
\begin{equation}
\Gamma _{64}^1=-\frac 1cB_1=\Gamma _{52}^3.\smallskip   \label{nuev}
\end{equation}

\end{document}